\documentclass[amsmath,twocolumn,prb,aps]{revtex4-1}
\usepackage{amsthm,amsfonts,graphicx,verbatim, color}
\usepackage{bm}
\usepackage[utf8]{inputenc}

\newcommand{\bcon}{e^{i\Phi_\Gamma}}

\newcommand{\ph}{\mathcal{PH}}

\newcommand{\dconf}{$\{ d \}$ }
\newcommand{\weier}{f}
\newcommand{\ftheta}{\theta}

\newcommand{\beqn}{\begin{eqnarray}}
\newcommand{\eeqn}{\end{eqnarray}}

\begin{document}

\title{Berry phase and model wavefunction in the half-filled Landau Level}

\author{Scott D. Geraedts$^{1,2}$, Jie Wang$^2$, E. Rezayi$^3$ and F. D. M. Haldane$^2$}
\affiliation{$^1$Department of Electrical Engineering, Princeton University, Princeton NJ 08544, USA}
\affiliation{$^2$Department of Physics, Princeton University, Princeton NJ 08544, USA }
\affiliation{$^3$Department of Physics, California State University, Los Angeles, CA 90032, USA}

\begin{abstract}
We construct model wavefunctions for the half-filled Landau level parameterized by 
``composite fermion occupation-number configurations'' in a two-dimensional
momentum space which correspond to a Fermi sea with particle-hole excitations.
When these correspond to a weakly-excited Fermi sea, they
have large overlap  with wavefunctions obtained by exact diagonalization
of lowest-Landau-level electrons interacting with a Coulomb interaction, allowing exact states to be identified with quasiparticle configurations.
We then formulate a many-body version of the single-particle Berry phase for
adiabatic transport of a single quasiparticle around a path in momentum space, and evaluate it using a sequence of exact eigenstates in which a single quasiparticle
moves incrementally.    In this formulation the standard free-particle construction in terms of the overlap between ``periodic parts of successive Bloch wavefunctions'' is reinterpreted as the matrix element of a ``momentum boost'' operator between the full Bloch states, which becomes the matrix elements of a Girvin-MacDonald-Platzman
density operator in the many-body context.
This allows computation of the Berry phase for transport of  a single composite fermion around the Fermi surface.  In addition to a phase contributed by the density operator, we find a phase of exactly $\pi$ for this process. 
\end{abstract}
\maketitle

\section{Introduction}
Two dimensional electron gases in high magnetic fields exhibit a wide variety of interesting physical properties. Perhaps most notable of these is the quantum Hall effect, which is a classic example of a topological phase. Another interesting phase occurs at even-denominator filling, the so-called ``composite Fermi liquid'' (CFL).\cite{HLR} This compressible phase is traditionally thought of as a Fermi liquid of ``composite fermions''\cite{JainCF89,LopezFradkin91}--bound states of electrons and even numbers of flux quanta, which experience no net magnetic field.

The composite Fermi liquid has been described theoretically in a number of complementary ways, such as through a model wavefunction\cite{HaldaneRezayi}, an effective field theory, called Halperin-Lee-Read (HLR) theory\cite{HLR} and through flux attachment\cite{JainCF89}. 
When projected into a single Landau level, the problem at half-filling ($\nu=1/2$) has a particle-hole symmetry which takes $\nu \rightarrow 1-\nu$. It is unclear how the various descriptions realize this particle-hole symmetry. At small sizes the model wavefunction has numerically been found to be very close to particle hole symmetric. 
It is difficult to see how the other descriptions behave under particle-hole symmetry because in order for this symmetry to exist we must project into a single Landau level, an analytically difficult procedure. Before this projection, the descriptions are clearly not particle-hole symmetric, an issue which has been discussed in a number of previous works\cite{Kivelson1997,Lee1998,PasquierHaldane1998,Read1998,MurthyShankarRMP}. 

Recently Son\cite{Son} has proposed an alternative to the HLR theory which is particle-hole symmetric even before Landau level projection. 
 In Son's theory the composite fermions are neutral Dirac fermions. One consequence of this is that the composite fermions should acquire a Berry phase of $\pi$ when moved around the Fermi surface.  This $\pi$ Berry phase has been indirectly confirmed numerically by observing an absence of $\pi$-backscattering in a density matrix renormalization group study,~\cite{Geraedts} but only for composite fermions that sit exactly on the Fermi surface. A direct measurement of the Berry phase for a wider variety of paths in momentum space is the main result of this work.


The Berry phase factor  $\exp (i \Phi_{\Gamma})$ is the phase  acquired when a quantum state is adiabatically evolved  around a closed path $\Gamma$  in parameter space, while remaining in the same Hilbert space.   When a single-particle Bloch state is evolved around a path in momentum space, this must be modified, as Bloch states with different Bloch vectors $\bm k$ belong to different Hilbert subspaces, and cannot be compared.   The usual approach is to factorize the Bloch wavefunction $\Psi_{\bm k}(\bm x)$  into a periodic part  $u_{\bm k}( \bm x)$ times a Bloch factor $\exp (i \bm k\cdot \bm x)$, and treat the periodic factor alone as the ``wavefunction''.     We 
can reinterpret the overlap $\langle u_{\bm k_1}|u_{\bm k_2}\rangle$ as the matrix element
$\langle \psi_{\bm k_1} |\rho (\bm k_1 - \bm k_2)|\psi_{\bm k_2}\rangle$, where $\bm \rho(\bm q)$ is the Fourier-transformed density operator $\exp (i \bm q\cdot \bm r)$.

We generalize the momentum space Berry phase to a many-body system in the following way:
\begin{eqnarray}
&&\bcon = \prod_{{\rm path} }  \langle \Psi(\{\bm k_i'\} )| \rho (\bm q) | \Psi(\{\bm k_i\} ) \rangle
\label{berrydef2} \\
&&\rho(\bm q)=\sum_{i=1}^N  e^{i \bm q \cdot \bm r_i},
\quad \bm k_i' = \bm k_i + \bm q\delta_{i,N}.
 \label{rhodef}
\end{eqnarray}
In the above equation we have assumed that $\Psi$ can be expressed as a Slater determinant of $N$ composite fermions, each with momentum $\bm k_i$. The ``path'' is a closed path in momentum space taken by  $\bm k_N$,  the momentum  of the composite fermion which moves around the path, while the others are unchanged.

In this paper we calculate the Berry phase from Eq.~(\ref{berrydef2}) using exact diagonalization (ED) in the basis of ``guiding center states'' left after projection into a single Landau level.   This also means that the operators $\rho(\bm q)$ becomes the non-commutative Girvin-MacDonald-Platzman\cite{GMP} (GMP) operators.
To apply Eq.~(\ref{berrydef2}) we need to show that the states produced by ED can indeed be identified with the analogs of  Slater determinants of composite fermions, and assign a configuration  $\{\bm k\}$ to each ED state in the sequence.
We accomplish this by comparing the ED states to a model wavefunction \cite{Simons,MM_Scott,MM_Jie}. 


\section{Comparing a model wavefunction and numerically-obtained states}
\label{sec:wf}
The model wavefunction often  used for the composite Fermi liquid is a determinant of translation operators acting on a bosonic $\nu=1/2$ Laughlin state. When projected into the lowest Landau level, such a wavefunction has computational complexity $N!$ for $N$ electrons.
We consider instead the following model wavefunction\cite{JainKamilla,Shao,Jie_MonteCarlo}, which has only $N^3$ complexity. The wavefunction takes as variational parameters $\{ d \}$, which are the momenta of the composite fermions. 
\begin{eqnarray}
&&\Psi(\{z\},\{\alpha\},\{d\},\bar d)= \nonumber \\
&&\left[\prod_{k=1}^{m} \weier ( \sum_i z_i -\alpha_k) \right] \det{M} \left[ \prod_{i<j} \weier(z_i - z_j)^{m-2}\right] \label{wfdef}\\
&&M_{ij}=\exp\left(\frac{z_i d_j ^* - z_i ^* d_j}{2m}\right) \prod_{k\neq i} \weier(z_i - z_k - d_j + \bar d). \label{Mdef}
\end{eqnarray}
Here the $z_i$ are the electron positions, and $m=1/\nu$ (for the half-filled case $m=2$).
The function $f(z)=\sigma(z)e^{-\frac{1}{2mN}zz^*}$ is a modified version of the Weierstrass sigma function times Gaussian \cite{Jie_MonteCarlo}.
We are working on a torus with dimensions $L_1,L_2$ and $L_1^* L_2 - L_1 L_2^*=2\pi mNi$. 
\footnote{A downside to writing the wavefunction in terms of the $f(z)$ is that it does not look holomorphic. However, the factor of $\exp(-\frac{1}{2mN}zz^*)$ contained in the definition of $f(z)$ does cancel the $z^*$ which appear in this equation}.

The first term in Eq.~(\ref{wfdef}) is the usual center-of-mass term present in model wavefunctions on the torus. The parameters $\alpha_k$ set the basis of the $m$-fold degenerate ground states, and by translational symmetry they must satisfy the constraint
\begin{equation}
\sum_k \alpha_k = N \bar d,
\end{equation}
and $\bar d$ is another variational parameter which we choose to be equal to the average value of the $\{ d \}$.

The determinant can be interpreted as attaching each electron to an area of reduced density. 
The object which results from this flux attachment is neutral and has a dipole moment $d$ which is a free parameter of the wavefunction.
We interpret this object as a composite fermion, and by the commutation relations of a single Landau level, the dipole moment specifies the momentum of this composite fermion, $(d_x,d_y)=(k_y,-k_x)$. In order for the wavefunction to satisfy translational symmetry, both the $d_x$ and $d_y$ must take values $\mathbb{Z}/N$ \cite{Jie_MonteCarlo}. The final term in Eq.~(\ref{wfdef}) is what is left of the Vandermonde factor after two powers of $(z_i-z_j)$ were used to construct the determinant.

We now wish to compare this model wavefunction to numerically obtained eigenstates expressed in second-quantized form.
To second-quantize the model wavefunction, consider that any many-body wave function $\Psi(\{z\})$ evaluated on a set of points $\{z\}$ can be written as:
\begin{equation}
\Psi(\{ z \})=\sum_{ \{n \} } c(\{n\}) \phi(\{z\},\{n\})
\end{equation}
where $\phi(\{z\},\{n\})$ is a Slater determinant of single-particle basis states for a single Landau level. The coefficients $c(\{n\})$ are of course independent of $\{z\}$, so by evaluating $\Psi(\{ z \})$ and $\phi(\{z\},\{n\})$ for many different sets of $\{z\}$ we can compute $c(\{n\})$ using a linear solver. This method relies on the fact that the wavefunction $\Psi(\{ z \})$ is already projected into a single Landau level, so that there are a finite number of single-particle basis states. Note that depending on the chosen sets of $\{z\}$, the matrix $ \phi(\{z\},\{n\})$ can be quite numerically unstable. Therefore we actually find it necessary to evaluate the wavefuntion at a number of $\{z\}$ points larger than the Hilbert space dimension, and to compute $c(\{n\})$ as the solution to a least-squares problem.

\begin{figure}
\includegraphics[width=\linewidth]{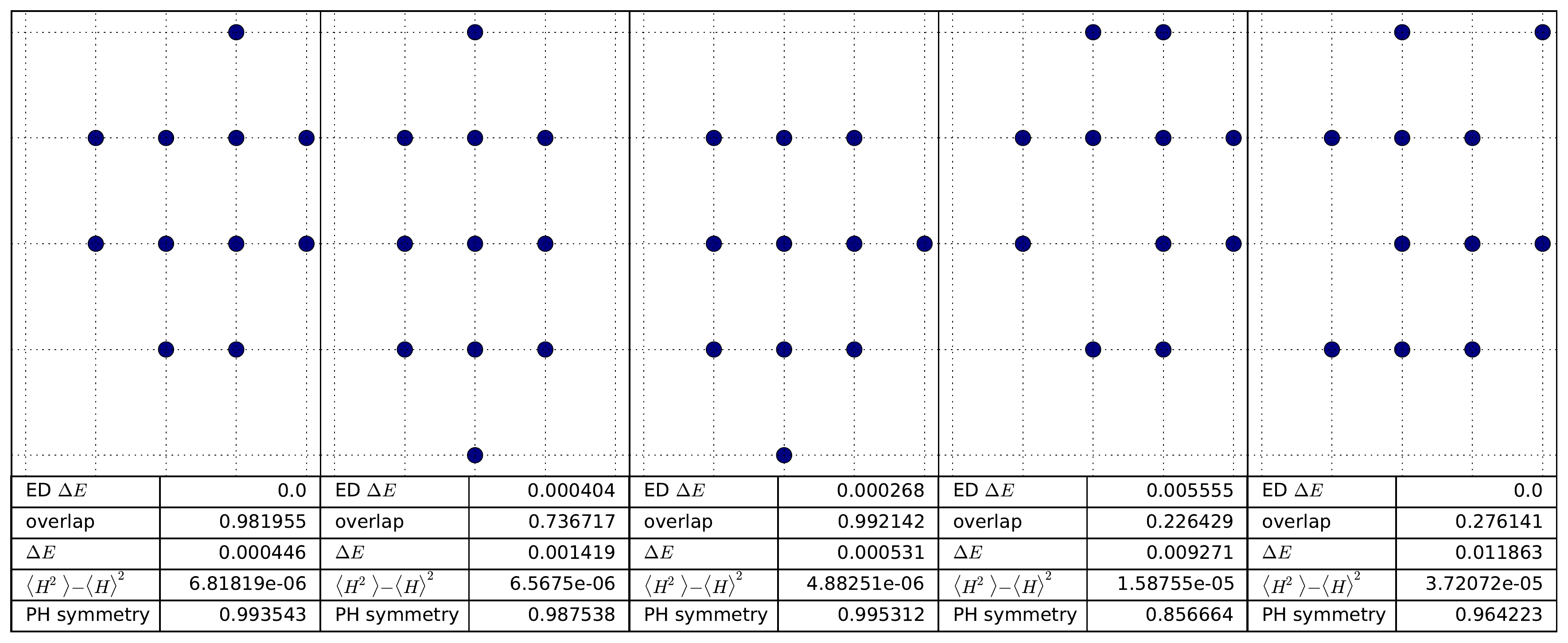}
\caption{ 
The energy, energy variance, overlap with exact eigenstates, and overlap with $\ph$ conjugates for a number of model states, with the $\{ d \}$ configurations shown. We see that clustered $\{ d \}$ configurations have low energies, good overlap with exact eigenstates, and are $\ph$ symmetric, while less clustered configurations lose these properties.
}
\label{FS_figure}
\end{figure}

Carrying out this procedure allows us to compare the model wavefunction to ED states. The results of this study are shown in Fig.~(\ref{FS_figure}), where we choose a number of different $\{ d \}$ configurations, and compute their energy, overlap with ED states and energy variance (which is zero for eigenstates of the Hamiltonian, and nonzero otherwise). 

Fig.~(\ref{FS_figure}) shows us that states where the $\{ d \}$ configurations are clustered are very close to the exact eigenstates. As the $\{ d \}$ become less clustered, these properties are reduced and also the energy increases. Our interpretation of these results is that the ground state and a few low-lying excited states can be well approximated by a composite fermion picture, but other states cannot. Note that Eq.~(\ref{wfdef}) is invariant under changing all the $\{ d \}$ by a fixed amount. This implies that the wavefunction is independent of the location of the center of the Fermi sea, it only depends on the shape of the Fermi sea.

One can try to use the model wavefunction to obtain multiple states for the same momentum $\bm  K$, by changing the \dconf configurations in a manner corresponding to exciting composite fermions out of the Fermi surface. However, we find that the different ``composite fermion excitations'' are not linearly independent. We can construct the model wavefunctions for all possible excitations which move composite fermion momenta by one lattice spacing in some direction, and compute the overlap matrix for the resulting states. We find that the rank of this matrix is always one less than the number of states. A possible physical explanation for this observation is that the there is an extra gauge degree of freedom that gaps out a Fermi surface excitation mode. This observation is analogous to the reduced central charge observed in Ref.~(\onlinecite{Geraedts}) and is due to the non-Fermi liquid character of the CFL.

\section{Particle-Hole Symmetry and Berry Phase}
\label{sec:berry}
Analytically evaluating the action of particle-hole symmetry on Eq.~(\ref{wfdef}) can in principle be done following Ref.~(\onlinecite{GirvinPH}), but is in practice intractable.
Once we have expressed the model wavefunction in the second-quantized basis, however, particle-hole symmetry is easy to implement since we just interchange the filled and empty states.  
For a model state with composite fermion dipole moments $\{ d \}$, we find that the overlap between its $\ph$ conjugate and the model state with dipole moments $\{-d \}$ is close to one, i.e. particle-hole symmetry flips the momenta of the composite fermions. 
Therefore the combination of $\ph$ and a rotation by $\pi$ ($R_\pi$) leaves the total momentum unchanged, so we can compute the quantity 
\begin{equation}
\langle \Psi| \ph~R_{\pi} |\Psi \rangle
\label{phrpi}
\end{equation}
and expect that it is $1$ for $\ph$ symmetric states. In Fig.~(\ref{FS_figure}) we give the values for this quantity for various $\{ d \}$ configurations, for both exact and model states. We see that the model states in which the composite fermions are clustered to form a Fermi sea are indeed nearly particle-hole symmetric. 

We now measure the Berry phase. For a given ED state, we use the overlap with the model wavefunction to determine whether that state can be well-described in terms of composite fermions, and also which composite fermions are filled for that state. We prepare a sequence states which consist of a filled ``Fermi surface'' plus or minus one composite fermion. We can compute Eq.~(\ref{berrydef2}) for a sequence of states where this ``extra'' composite fermion moves around some closed path. 
 
One subtlety when computing the Berry phase around the center of the Fermi sea is that the center of the Fermi sea is not gauge invariant: we can translate all the composite fermions by one lattice spacing without changing any physics. Therefore the notion of transporting a composite fermion ``around the center of the Fermi sea'' is only well-defined when a compact Fermi sea is present. Furthermore in order for a path around the Fermi surface to be defined we also have to exclude non-trivial paths around the torus. Therefore in what follows we must restrict ourself to the case where a composite Fermi sea exists, and the composite fermion we are moving is not too far from this Fermi sea (so that it cannot wind around the torus). The results of the previous section show that such states are the only states which are well described by a composite fermion picture.

\begin{figure}
\includegraphics[width=0.9\linewidth]{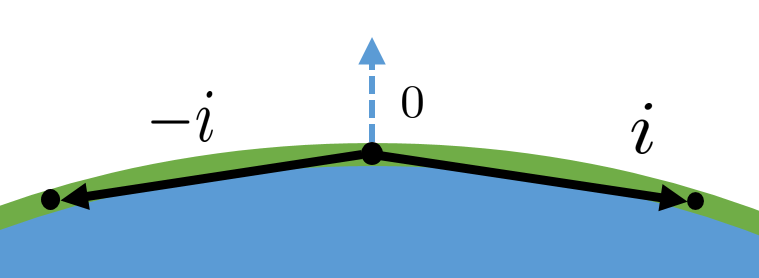}
\caption{The density operator in Eq.~(\ref{berrydef2}) adds an additional phase to our Berry phase calculation. As outlined in the text, particle-hole symmetry forces this phase to be imaginary while an antiunitary reflection symmetry (present in the thermodynamic limit, as well as in the square lattice considered here) introduces a relative $-1$ between clocckwise and counterclockwise hopping, while also forbidding hopping in a direction normal to the Fermi surface.}
\label{extraphase}
\end{figure}

Another issue when computing the ``many-body Berry-like  phase'' defined in Eq.~(\ref{berrydef2}) is that we expect that the phase  should behave smoothy in the thermodynamic limit as the individual steps in the path become infinitesimal and the the path in momentum space becomes smooth and continuous.   However we did observe
an additional geometric phase factor associated with the 
the projected GMP density operator $\rho(\bm q)$ in each segment of the path.
Fortunately we can use the symmetries of the problem to determine what this phase is, and we summarize these results in Fig.~(\ref{extraphase}).
First consider the combination of particle hole and inversion symmetry discussed in Eq.~(\ref{phrpi}). This symmetry takes $\rho(q) \rightarrow -\rho(q)$ and $i \rightarrow -i$. A simple overlap behaves like  
$\langle \Psi_1| \Psi_2 \rangle \rightarrow \langle \Psi_1| \Psi_2 \rangle^*$ under this symmetry, while the matrix element  in Eq.~(\ref{berrydef2}) behaves like   
$\langle \Psi_1|\rho| \Psi_2 \rangle \rightarrow -\langle \Psi_1| \rho|\Psi_2 \rangle^*$. From this we deduce that the extra phase contributed by the $\rho(\bm q)$ is purely imaginary. 

Empirically, we observed that the expression for the phase factor defined by Eq. (\ref{berrydef2}) appears to be
\begin{equation}
e^{i\tilde \Phi_\Gamma} = (i)^{N_+-N_-} e^{i\tilde\Phi}
\label{rules}
\end{equation}
where $N_+$ are the number of discrete steps around the Fermi surface that are in the positve (anticlockwise ) sense and $N_-$ is the number in the negative (clockwise) sense, and $e^{i\tilde\Phi}$ is the ``true'' Berry phase factor which is  $(-1)^W$ if the path stays close to the Fermi surface with winding number $W$.

Another constraint on the phase associated with $\rho(\bm q)$ can be found from an antiunitary reflection symmetry about the (e.g.) $x$-axis, which takes $i\rightarrow -i$ and $k_x\rightarrow -k_x$. Such a symmetry exists for a torus with square or hexagonal boundary conditions, and it also exists in the thermodynamic limit. For a given composite fermion momentum $\bm k$, we can define an angle $\theta_k$ relative to the center of the Fermi sea. Such a definition requires that a compact Fermi sea exists, but we have seen in the previous section that this is true for composite Fermi liquid states. Given a composite fermion momentum $\bm k_i$ and a momentum change $\bm q$, we can then define $d\theta_q\equiv \theta_{k+q}-\theta_k$. The reflection symmetry takes $d\theta_q \rightarrow -d\theta_{q}$.  
It takes $\langle \Psi_1| \Psi_2 \rangle \rightarrow \langle \Psi_1| \Psi_2 \rangle^*$ and (since we already established that $\rho(\bm q)$ contributes a purely imaginary phase)
$\langle \Psi_1|\rho(d\theta_q)| \Psi_2 \rangle \rightarrow -\langle \Psi_1| \rho(-d\theta_{q})|\Psi_2 \rangle^*$. From this we know that the phase contributed by the $\rho(d\theta_q)$ changes sign when the sign of $d\theta_q$ changes sign.

Putting reflection and particle-hole symmetry together gives precisely the relation in Eq.~(\ref{rules}). 
Note that for odd $N_{\rm steps}$ one can change the sign of the Berry phase by going around the path $\Gamma$ in the opposite direction, to avoid dealing with this ambiguity we restrict to paths with even $N_{\rm steps}$. Also note that steps with $d\theta_q=0$ (perpendicular to the Fermi surface, see Fig.~\ref{extraphase}) are forbidden by the reflection symmetry.

\begin{figure}
\includegraphics[width=\linewidth]{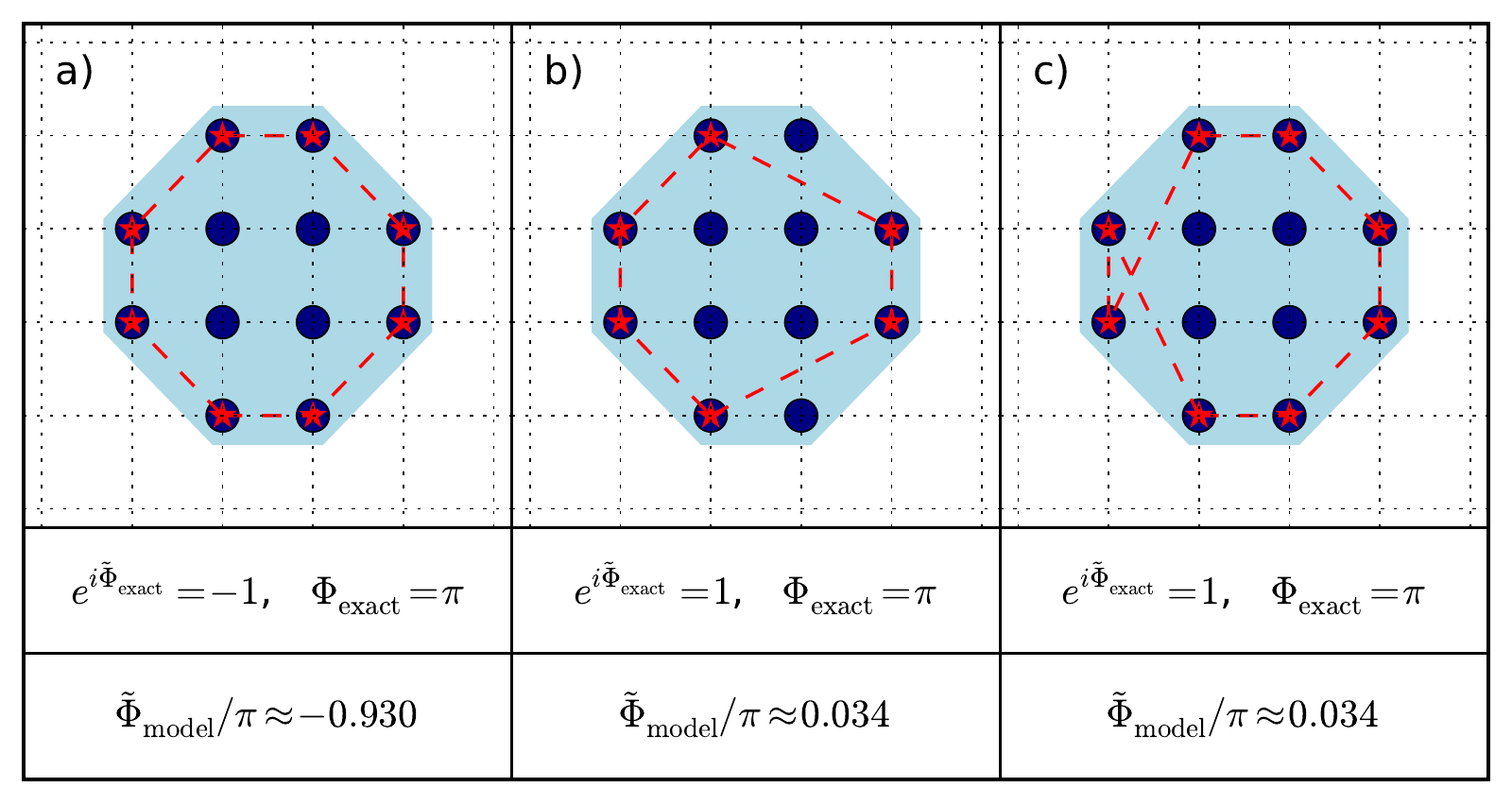}
\caption{Berry phases observed for a variety of paths around the Fermi surface. The blue circles represent the locations of composite fermions, and for each step on the path we remove a composite fermion at the location of the red stars (i.e. we are moving a composite hole around the Fermi surface). 
There are a number of effects which arise from the insertion of a density operator in Eq.~(\ref{berrydef2}). These differences lead to additional phases summarized in Eq.~(\ref{rules}).
We compute these Berry phase for the exact ED states in the first row of the figure, and for the model wavefunction of Eq.~(\ref{wfdef}) in the second row. In both cases the results, combined with Eq.~(\ref{rules}), indicate a Berry phase of $\pi$ when the center of the Fermi surface is encircled.
}
\label{berry_figure}
\end{figure}

The results of measuring the Berry phase can be seen in Fig.~(\ref{berry_figure}), where we show the $\tilde\Phi$ extracted from Eq.~(\ref{berrydef2}) for a variety of system sizes, and paths taken by the composite  fermion. In addition to computing these phases using exact wavefunction obtained from diagonalization of the Hamiltonian($\tilde\Phi_{\rm exact}$), we can compute them purely from the model wavefunction ($\tilde\Phi_{\rm model}$), providing another way of estimating how close to exact the model wavefunction is.
By comparing Figs.~(\ref{berry_figure})(a-c) we observe the sign structure predicted from the above symmetry analysis. Using Eq.~(\ref{rules}) we find that the true Berry phase $\tilde\Phi=\pi$.  We postulate that this our composite fermion encircled center of the Fermi surface. Note also that our Berry phase is always $0$ or $\pi$, there is no component related to the area enclosed by the path, consistent with the composite fermions seeing no external field.

\begin{figure}
\includegraphics[width=\linewidth]{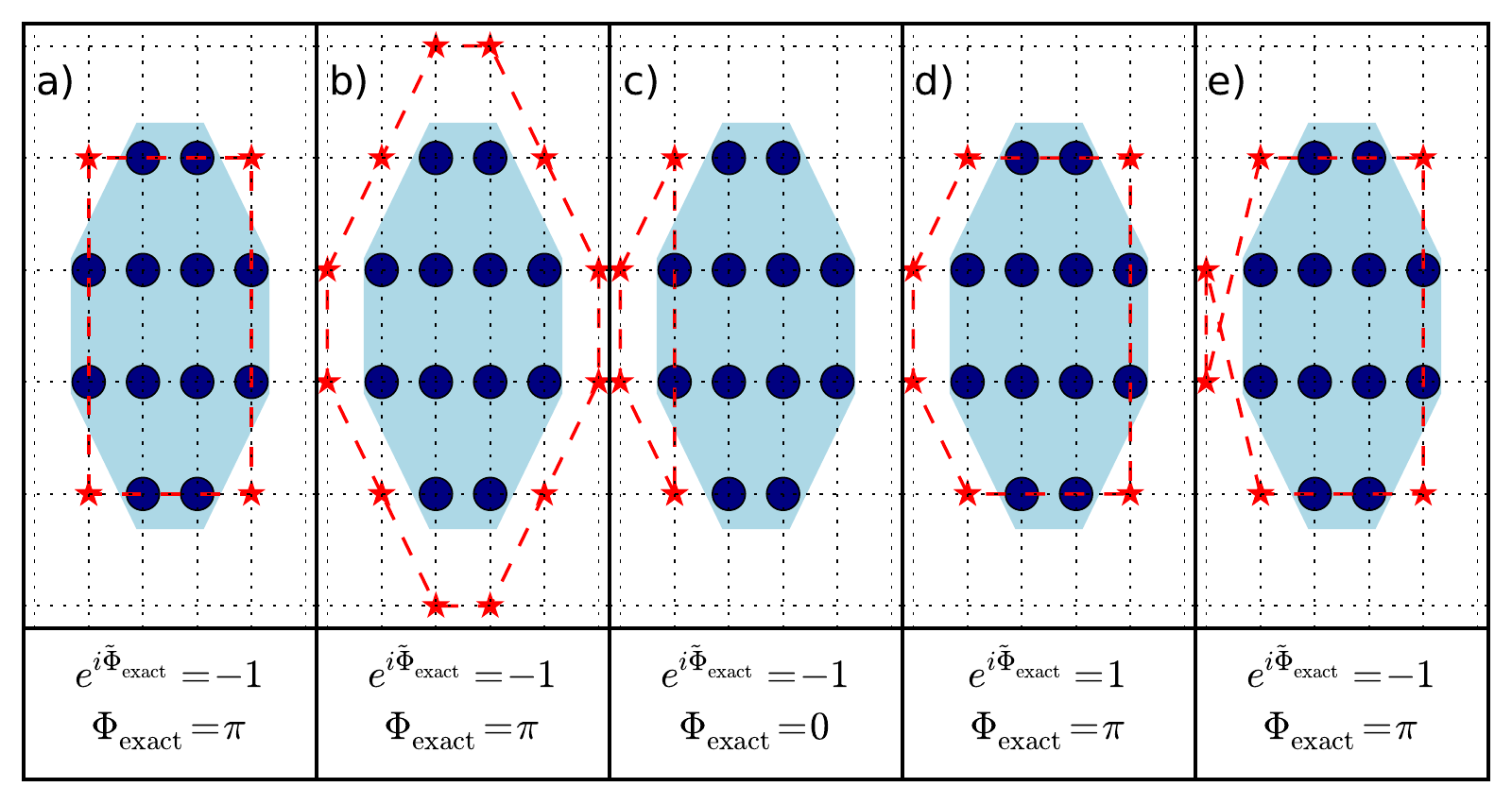}
\caption{Same as Fig.~\ref{berry_figure}, but for $N=13$. The red stars indicate an additional composite fermion which we are moving around the Fermi surface.
At this size we are able to construct paths which do not enclose the origin, and we find $\tilde\Phi_{\rm exact}=0$ for these paths.
}
\label{berry_figure13}
\end{figure}

In Fig.~(\ref{berry_figure13}) we perform the same analysis, this time with $N=13$. This system size is too large to second quantize the model wavefunction, and therefore we only compute $\tilde\Phi_{\rm exact}$ from the Hamiltonian. In our paper Ref.~(\onlinecite{Jie_MonteCarlo}), we describe an improved Monte Carlo procedure which enables us to compute the Berry phase $\tilde\Phi_{\rm model}$ of a model wavefunction for $N$ up to $70$. This supports the conclusions drawn in this work by allowing us to generate data like in Figs.~(\ref{berry_figure}) and (\ref{berry_figure13}) but for much larger sizes.

\section{ Discussion}
In this work we have used a model wavefunction for the half-filled Landau level to argue that the ground states obtained in exact diagonalization can be expressed as Slater determinants of non-interacting composite fermion states. We have shown that this description only holds when the composite fermion momenta are clustered into a Fermi surface-like configuration. These states are also particle-hole symmetric. We then computed the Berry phase upon taking a composite fermion around the Fermi surface. We show that this Berry phase is $\pi$ when the path taken by the composite fermion encloses the Fermi surface, and zero otherwise. It is also interesting to test how close to exact for other candidate model wavefunctions (see Appendix A).

Our results are consistent with the theory of Son\cite{Son}, in which the Berry phase arises from the Dirac nature of the composite fermions. However there are other aspects of this effective theory which cannot be seen in our numerics. 
The composite fermions discussed in our work are single-component objects, and the relation to a composite Dirac fermion is unknown. Additionally, we have found the model wavefunction only applies near the Fermi surface, so it cannot be used to attempt to observe a Dirac-like dispersion relation deep in the Fermi sea. In fact, for a system with inversion symmetry the ground state has exactly zero overlap with its particle-hole conjugate at the center of the Fermi surface, implying that in our microscopic model it is meaningless to discuss a Dirac singularity.

We recently become aware of Ref.~(\onlinecite{Simons}) by M. Fremling et.al. where the authors performed similar analysis of CFL's energy, particle-hole symmetry and overlap property \cite{MM_Scott}, by a different lowest Landau level projection method. To compute the Berry phase, M. Fremling et.al. moved two composite fermions at opposite sides of the Fermi disc for $N$ steps and found a phase of $e^{i\pi\cdot (N-1)}$, which is consistent with the rule in Eqn.~(\ref{rules}) and described in Ref.~(\onlinecite{MM_Scott,MM_Jie,Jie_MonteCarlo}) by taking $N_+=2N$, $N_-=0$, $\tilde\Phi=\pi$.


\appendix
\section{Comparison of different CFL model wavefunctions}
\label{sec:wfcompare}

The wavefunction in Eq.~(\ref{wfdef}) is only one possible model wavefunction for the composite Fermi liquid. Here we compare its overlap with exact states and particle-hole symmetry to two other candidate wavefunctions.

The first wavefunction we consider is the traditional Halperin-Lee-Read inspired wavefunction\cite{HLR,JainCF89}. Schematically this can be written as follows (in the lowest Landau level):
\begin{equation}
\det[e^{\frac12 (z_id_j^* -z_i^*d_j) }] \prod_{i<j} (z_i-z_j)^2 
\label{HLRschematic}
\end{equation}
To project this to the lowest Landau level we need to replace all the $z_i^*$ with translation operators. To do this we need to expand the determinant as a sum over all permutations of the indices, and replace $d_j$ with $d_{P(i)}$, where $P(i)$ represents the $i^{th}$ index of the permutation. The projected wavefunction on the torus is\cite{Shao}:
\begin{eqnarray}
&&\left[ \sum_{P}^{N_e!} (-1)^P M_P \right] \sum_{k=1}^2 f(Z-\alpha_k) \label{HLRfull}\\
&& M_P \equiv \prod_i e^{\frac12 (z_id_j^* -z_i^*d_j)} \prod_{i<j} f(z_i-z_j-d_{P(i)}+d_{P(j)})^2. \nonumber
\end{eqnarray}
Similar to \ref{wfdef}, though this wavefunction does not look holomorphic the Gaussian factors in $f(z)$ cancel the $z_i^*$ parts. Compared to Eq.~(\ref{HLRschematic}) we have written this wavefunction on a torus by using the elliptic function $f(z)$ and by adding the center-of-mass part which is a function of $Z$. Unlike Eq.~(\ref{wfdef}), the $d_j$ in this wavefunction are defined on a lattice of spacing $1/N_{\phi}$, not $1/N_e$. The quantization of the $d_j$ is set by the condition that the boundary conditions of the wavefunction are independent of the positions of the $z_i$,  one can show using the translation properties of the $f(z)$ that for this wavefunction a spacing of $1/N_\phi$ is sufficient to guarantee this. 
For periodic boundary conditions the sum of the $\alpha_k$ must be equal to {\it twice} the sum of the $d_j$, this can also be seen from the translation properties of the $f(z)$.
The wavefunction in Eq.~(\ref{wfdef}) is an approximation to Eq.~(\ref{HLRfull}), which was obtained by taking the Jastrow factors in Eq.~(\ref{HLRschematic}) inside the determinant. The reason for this is since Eq~(\ref{wfdef}) is a total determinant it can be evaluated in polynomial time (while the sum over $P(i)$ in Eq.~(\ref{HLRfull}) takes exponential time). This significantly speeds up numerical calculations, such as the comparisons to exact states in the previous sections as well as Monte Carlo calculations.

We can think of Eq.~(\ref{HLRschematic}) as inserting two flux quanta a distance $d_j$ away from each electron, creating a composite object with dipole moment $2e$. 
Wang and Senthil\cite{WangSenthil} have suggested that one could instead put one flux quantum on the electron, and one displaced by $d_j$. Schematically, a wavefunction capturing this picture looks like:
\begin{equation}
\prod_{i<j} (z_i-z_j)  \det[e^{\frac14 (z_id_j^* -z_i^*d_j) }] \prod_{i<j} (z_i-z_j).
\label{senthilSchematic}
\end{equation}
Expanding the determinant as above leads to:
\begin{eqnarray}
\left[ \sum_{P}^{N_e!} (-1)^P M_P \right] &&\sum_{k=1}^2 f(Z-\alpha_k)\prod_{j>i}f(z_i-z_j) \\
M_P \equiv &&\prod_i e^{\frac14 (z_id_{P(i)}^* -z_i^*d_{P(i)})} \times \nonumber\\
&&\prod_{i<j} f(z_i-z_j-d_{P(i)}+d_{P(j)}). \nonumber
\label{senthilFull}
\end{eqnarray}
Here, as in Eq.~(\ref{wfdef}) but unlike Eq.~(\ref{HLRfull}) the $d_j$ live on a $1/N_e$ lattice and the system has periodic boundary conditions when the sum of the $\alpha_k$ is equal to the sum of the $d_j$. This wavefunction is inspired by the composite fermion structure postulated by Son, and it is interesting to ask whether it is therefore more particle-hole symmetric than Eq.~(\ref{HLRfull}).

\begin{figure}
\includegraphics[width=0.9\linewidth]{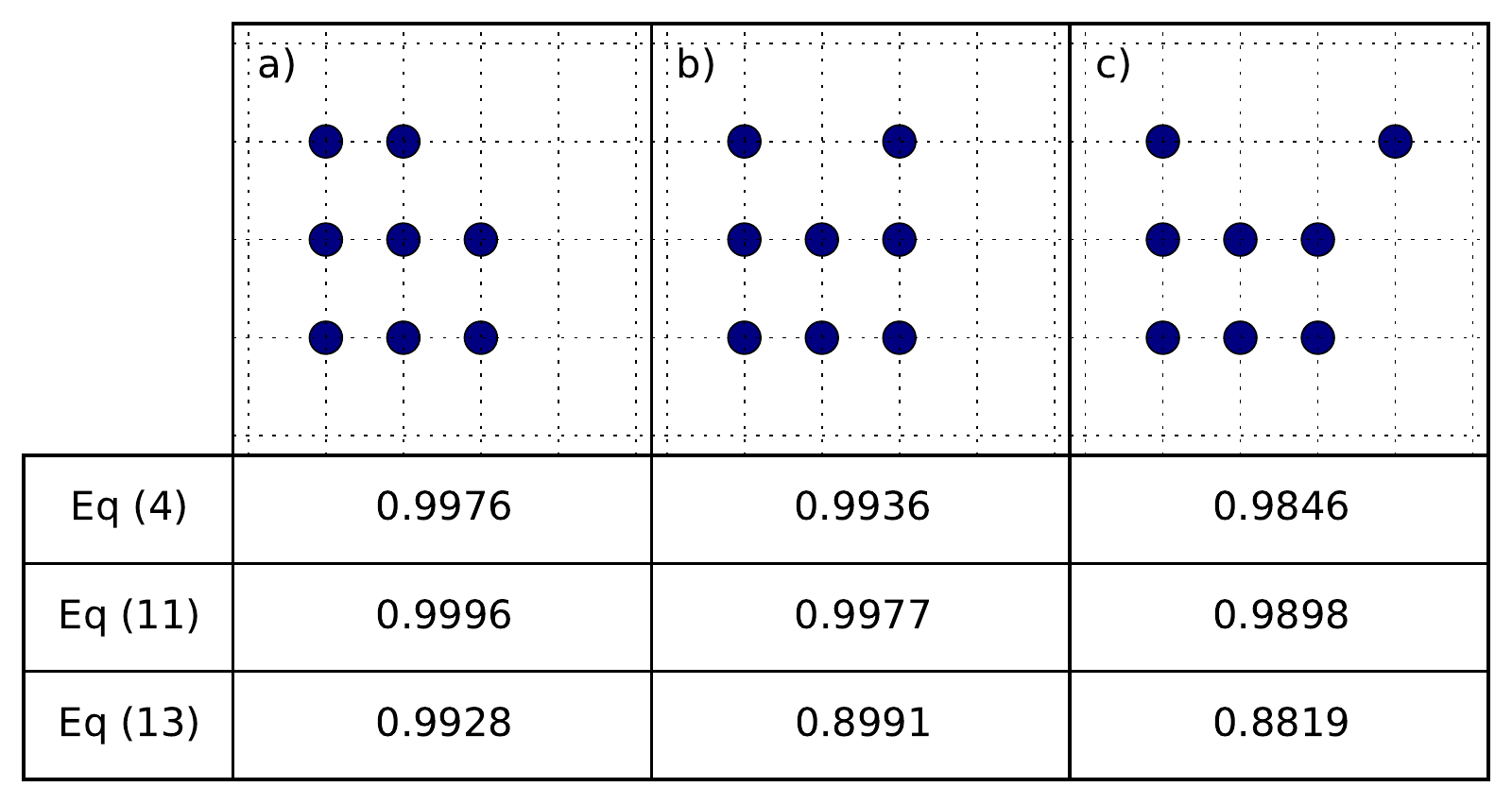}
\caption{ The overlaps between a wavefunction and its particle-hole conjugate, for three different configurations of composite fermions (shown in the top row) and three different wavefunctions: the ``determinant'' wavefunction from Eq.~(\ref{wfdef}), the standard HLR wavefunction from Eq.~(\ref{HLRfull}) and the wavefunction inspired by Ref.~(\onlinecite{WangSenthil}) where the composite fermion is a $+e/2:-e/2$ dipole from Eq.~(\ref{senthilFull}). We see that the standard HLR wavefunction produces the most particle-hole symmetric wavefunctions, while the modifed version produces the least symmetric ones.  }
\label{comparefig}
\end{figure}

Fig.~(\ref{comparefig}) shows our results. For a number of different configurations of composite fermions, we have evaluated the particle-hole symmetry by taking the overlap between a model wavefunction and its particle-hole conjugate. Compared to the previous sections, we are limited to smaller system sizes since the wavefunctions discussed in this section are slower to evaluate. Since the exact states are particle-hole symmetric this measurement also probes how similar the model wavefunctions are to the exact states.
Our results show that the HLR wavefunction is the most particle-hole symmetric. They also show that he approximate wavefunction Eq.~(\ref{wfdef}) is only slightly less symmetric, while the wavefunction of Eq.~(\ref{senthilFull}) is the least symmetric. The similarity between the determinant wavefunction and the standard HLR wavefunction gives us confidence that we can use the determinant wavefunction, which is computationally much more efficient, to study the properties of the CFL. All three of the wavefunctions discussed in this section have the same Berry phase properties.

\acknowledgments

This work was supported by Department of Energy BES Grant DE-SC0002140.

\bibliography{cfl}

\begin{thebibliography}{21}%
\makeatletter
\providecommand \@ifxundefined [1]{%
 \@ifx{#1\undefined}
}%
\providecommand \@ifnum [1]{%
 \ifnum #1\expandafter \@firstoftwo
 \else \expandafter \@secondoftwo
 \fi
}%
\providecommand \@ifx [1]{%
 \ifx #1\expandafter \@firstoftwo
 \else \expandafter \@secondoftwo
 \fi
}%
\providecommand \natexlab [1]{#1}%
\providecommand \enquote  [1]{``#1''}%
\providecommand \bibnamefont  [1]{#1}%
\providecommand \bibfnamefont [1]{#1}%
\providecommand \citenamefont [1]{#1}%
\providecommand \href@noop [0]{\@secondoftwo}%
\providecommand \href [0]{\begingroup \@sanitize@url \@href}%
\providecommand \@href[1]{\@@startlink{#1}\@@href}%
\providecommand \@@href[1]{\endgroup#1\@@endlink}%
\providecommand \@sanitize@url [0]{\catcode `\\12\catcode `\$12\catcode
  `\&12\catcode `\#12\catcode `\^12\catcode `\_12\catcode `\%12\relax}%
\providecommand \@@startlink[1]{}%
\providecommand \@@endlink[0]{}%
\providecommand \url  [0]{\begingroup\@sanitize@url \@url }%
\providecommand \@url [1]{\endgroup\@href {#1}{\urlprefix }}%
\providecommand \urlprefix  [0]{URL }%
\providecommand \Eprint [0]{\href }%
\providecommand \doibase [0]{http://dx.doi.org/}%
\providecommand \selectlanguage [0]{\@gobble}%
\providecommand \bibinfo  [0]{\@secondoftwo}%
\providecommand \bibfield  [0]{\@secondoftwo}%
\providecommand \translation [1]{[#1]}%
\providecommand \BibitemOpen [0]{}%
\providecommand \bibitemStop [0]{}%
\providecommand \bibitemNoStop [0]{.\EOS\space}%
\providecommand \EOS [0]{\spacefactor3000\relax}%
\providecommand \BibitemShut  [1]{\csname bibitem#1\endcsname}%
\let\auto@bib@innerbib\@empty
\bibitem [{\citenamefont {Halperin}\ \emph {et~al.}(1993)\citenamefont
  {Halperin}, \citenamefont {Lee},\ and\ \citenamefont {Read}}]{HLR}%
  \BibitemOpen
  \bibfield  {author} {\bibinfo {author} {\bibfnamefont {B.~I.}\ \bibnamefont
  {Halperin}}, \bibinfo {author} {\bibfnamefont {P.~A.}\ \bibnamefont {Lee}}, \
  and\ \bibinfo {author} {\bibfnamefont {N.}~\bibnamefont {Read}},\ }\href
  {\doibase 10.1103/PhysRevB.47.7312} {\bibfield  {journal} {\bibinfo
  {journal} {Phys. Rev. B}\ }\textbf {\bibinfo {volume} {47}},\ \bibinfo
  {pages} {7312} (\bibinfo {year} {1993})}\BibitemShut {NoStop}%
\bibitem [{\citenamefont {Jain}(1989)}]{JainCF89}%
  \BibitemOpen
  \bibfield  {author} {\bibinfo {author} {\bibfnamefont {J.~K.}\ \bibnamefont
  {Jain}},\ }\href {\doibase 10.1103/PhysRevLett.63.199} {\bibfield  {journal}
  {\bibinfo  {journal} {Phys. Rev. Lett.}\ }\textbf {\bibinfo {volume} {63}},\
  \bibinfo {pages} {199} (\bibinfo {year} {1989})}\BibitemShut {NoStop}%
\bibitem [{\citenamefont {Lopez}\ and\ \citenamefont
  {Fradkin}(1991)}]{LopezFradkin91}%
  \BibitemOpen
  \bibfield  {author} {\bibinfo {author} {\bibfnamefont {A.}~\bibnamefont
  {Lopez}}\ and\ \bibinfo {author} {\bibfnamefont {E.}~\bibnamefont
  {Fradkin}},\ }\href {\doibase 10.1103/PhysRevB.44.5246} {\bibfield  {journal}
  {\bibinfo  {journal} {Phys. Rev. B}\ }\textbf {\bibinfo {volume} {44}},\
  \bibinfo {pages} {5246} (\bibinfo {year} {1991})}\BibitemShut {NoStop}%
\bibitem [{\citenamefont {Rezayi}\ and\ \citenamefont
  {Haldane}(2000)}]{HaldaneRezayi}%
  \BibitemOpen
  \bibfield  {author} {\bibinfo {author} {\bibfnamefont {E.~H.}\ \bibnamefont
  {Rezayi}}\ and\ \bibinfo {author} {\bibfnamefont {F.~D.~M.}\ \bibnamefont
  {Haldane}},\ }\href {\doibase 10.1103/PhysRevLett.84.4685} {\bibfield
  {journal} {\bibinfo  {journal} {Phys. Rev. Lett.}\ }\textbf {\bibinfo
  {volume} {84}},\ \bibinfo {pages} {4685} (\bibinfo {year}
  {2000})}\BibitemShut {NoStop}%
\bibitem [{\citenamefont {Kivelson}\ \emph {et~al.}(1997)\citenamefont
  {Kivelson}, \citenamefont {Lee}, \citenamefont {Krotov},\ and\ \citenamefont
  {Gan}}]{Kivelson1997}%
  \BibitemOpen
  \bibfield  {author} {\bibinfo {author} {\bibfnamefont {S.~A.}\ \bibnamefont
  {Kivelson}}, \bibinfo {author} {\bibfnamefont {D.-H.}\ \bibnamefont {Lee}},
  \bibinfo {author} {\bibfnamefont {Y.}~\bibnamefont {Krotov}}, \ and\ \bibinfo
  {author} {\bibfnamefont {J.}~\bibnamefont {Gan}},\ }\href {\doibase
  10.1103/PhysRevB.55.15552} {\bibfield  {journal} {\bibinfo  {journal} {Phys.
  Rev. B}\ }\textbf {\bibinfo {volume} {55}},\ \bibinfo {pages} {15552}
  (\bibinfo {year} {1997})}\BibitemShut {NoStop}%
\bibitem [{\citenamefont {Lee}(1998)}]{Lee1998}%
  \BibitemOpen
  \bibfield  {author} {\bibinfo {author} {\bibfnamefont {D.-H.}\ \bibnamefont
  {Lee}},\ }\href {\doibase 10.1103/PhysRevLett.80.4745} {\bibfield  {journal}
  {\bibinfo  {journal} {Phys. Rev. Lett.}\ }\textbf {\bibinfo {volume} {80}},\
  \bibinfo {pages} {4745} (\bibinfo {year} {1998})}\BibitemShut {NoStop}%
\bibitem [{\citenamefont {Pasquier}\ and\ \citenamefont
  {Haldane}(1998)}]{PasquierHaldane1998}%
  \BibitemOpen
  \bibfield  {author} {\bibinfo {author} {\bibfnamefont {V.}~\bibnamefont
  {Pasquier}}\ and\ \bibinfo {author} {\bibfnamefont {F.~D.~M.}\ \bibnamefont
  {Haldane}},\ }\href {\doibase 10.1016/S0550-3213(98)00069-8} {\bibfield
  {journal} {\bibinfo  {journal} {Nucl. Phys. B}\ }\textbf {\bibinfo {volume}
  {516}},\ \bibinfo {pages} {719 } (\bibinfo {year} {1998})}\BibitemShut
  {NoStop}%
\bibitem [{\citenamefont {Read}(1998)}]{Read1998}%
  \BibitemOpen
  \bibfield  {author} {\bibinfo {author} {\bibfnamefont {N.}~\bibnamefont
  {Read}},\ }\href {\doibase 10.1103/PhysRevB.58.16262} {\bibfield  {journal}
  {\bibinfo  {journal} {Phys. Rev. B}\ }\textbf {\bibinfo {volume} {58}},\
  \bibinfo {pages} {16262} (\bibinfo {year} {1998})}\BibitemShut {NoStop}%
\bibitem [{\citenamefont {Murthy}\ and\ \citenamefont
  {Shankar}(2003)}]{MurthyShankarRMP}%
  \BibitemOpen
  \bibfield  {author} {\bibinfo {author} {\bibfnamefont {G.}~\bibnamefont
  {Murthy}}\ and\ \bibinfo {author} {\bibfnamefont {R.}~\bibnamefont
  {Shankar}},\ }\href {\doibase 10.1103/RevModPhys.75.1101} {\bibfield
  {journal} {\bibinfo  {journal} {Rev. Mod. Phys.}\ }\textbf {\bibinfo {volume}
  {75}},\ \bibinfo {pages} {1101} (\bibinfo {year} {2003})}\BibitemShut
  {NoStop}%
\bibitem [{\citenamefont {Son}(2015)}]{Son}%
  \BibitemOpen
  \bibfield  {author} {\bibinfo {author} {\bibfnamefont {D.~T.}\ \bibnamefont
  {Son}},\ }\href {\doibase 10.1103/PhysRevX.5.031027} {\bibfield  {journal}
  {\bibinfo  {journal} {Phys. Rev. X}\ }\textbf {\bibinfo {volume} {5}},\
  \bibinfo {pages} {031027} (\bibinfo {year} {2015})}\BibitemShut {NoStop}%
\bibitem [{\citenamefont {Geraedts}\ \emph {et~al.}(2016)\citenamefont
  {Geraedts}, \citenamefont {Zaletel}, \citenamefont {Mong}, \citenamefont
  {Metlitski}, \citenamefont {Vishwanath},\ and\ \citenamefont
  {Motrunich}}]{Geraedts}%
  \BibitemOpen
  \bibfield  {author} {\bibinfo {author} {\bibfnamefont {S.}~\bibnamefont
  {Geraedts}}, \bibinfo {author} {\bibfnamefont {M.}~\bibnamefont {Zaletel}},
  \bibinfo {author} {\bibfnamefont {R.}~\bibnamefont {Mong}}, \bibinfo {author}
  {\bibfnamefont {M.}~\bibnamefont {Metlitski}}, \bibinfo {author}
  {\bibfnamefont {A.}~\bibnamefont {Vishwanath}}, \ and\ \bibinfo {author}
  {\bibfnamefont {O.}~\bibnamefont {Motrunich}},\ }\href@noop {} {\bibfield
  {journal} {\bibinfo  {journal} {Science}\ }\textbf {\bibinfo {volume}
  {352}},\ \bibinfo {pages} {197} (\bibinfo {year} {2016})}\BibitemShut
  {NoStop}%
\bibitem [{\citenamefont {Girvin}\ \emph {et~al.}(1985)\citenamefont {Girvin},
  \citenamefont {MacDonald},\ and\ \citenamefont {Platzman}}]{GMP}%
  \BibitemOpen
  \bibfield  {author} {\bibinfo {author} {\bibfnamefont {S.~M.}\ \bibnamefont
  {Girvin}}, \bibinfo {author} {\bibfnamefont {A.~H.}\ \bibnamefont
  {MacDonald}}, \ and\ \bibinfo {author} {\bibfnamefont {P.~M.}\ \bibnamefont
  {Platzman}},\ }\href {\doibase 10.1103/PhysRevLett.54.581} {\bibfield
  {journal} {\bibinfo  {journal} {Phys. Rev. Lett.}\ }\textbf {\bibinfo
  {volume} {54}},\ \bibinfo {pages} {581} (\bibinfo {year} {1985})}\BibitemShut
  {NoStop}%
\bibitem [{\citenamefont {Fremling}\ \emph {et~al.}(2017)\citenamefont
  {Fremling}, \citenamefont {Moran}, \citenamefont {Slingerland},\ and\
  \citenamefont {Simon}}]{Simons}%
  \BibitemOpen
  \bibfield  {author} {\bibinfo {author} {\bibfnamefont {M.}~\bibnamefont
  {Fremling}}, \bibinfo {author} {\bibfnamefont {N.}~\bibnamefont {Moran}},
  \bibinfo {author} {\bibfnamefont {J.~K.}\ \bibnamefont {Slingerland}}, \ and\
  \bibinfo {author} {\bibfnamefont {S.~H.}\ \bibnamefont {Simon}},\ }\href@noop
  {} {\bibfield  {journal} {\bibinfo  {journal} {arXiv:cond-mat}\ ,\ \bibinfo
  {pages} {1711.01217}} (\bibinfo {year} {2017})}\BibitemShut {NoStop}%
\bibitem [{\citenamefont {Geraedts}\ \emph {et~al.}(2017)\citenamefont
  {Geraedts}, \citenamefont {Wang},\ and\ \citenamefont {Haldane}}]{MM_Scott}%
  \BibitemOpen
  \bibfield  {author} {\bibinfo {author} {\bibfnamefont {S.}~\bibnamefont
  {Geraedts}}, \bibinfo {author} {\bibfnamefont {J.}~\bibnamefont {Wang}}, \
  and\ \bibinfo {author} {\bibfnamefont {F.~D.~M.}\ \bibnamefont {Haldane}},\
  }\href {http://meetings.aps.org/link/BAPS.2017.MAR.K27.3} {\bibfield
  {journal} {\bibinfo  {journal}
  {http://meetings.aps.org/link/BAPS.2017.MAR.K27.3}\ } (\bibinfo {year}
  {2017})}\BibitemShut {NoStop}%
\bibitem [{\citenamefont {Wang}\ \emph
  {et~al.}(2017{\natexlab{a}})\citenamefont {Wang}, \citenamefont {Geraedts},
  \citenamefont {Rezayi},\ and\ \citenamefont {Haldane}}]{MM_Jie}%
  \BibitemOpen
  \bibfield  {author} {\bibinfo {author} {\bibfnamefont {J.}~\bibnamefont
  {Wang}}, \bibinfo {author} {\bibfnamefont {S.}~\bibnamefont {Geraedts}},
  \bibinfo {author} {\bibfnamefont {E.~H.}\ \bibnamefont {Rezayi}}, \ and\
  \bibinfo {author} {\bibfnamefont {F.~D.~M.}\ \bibnamefont {Haldane}},\ }\href
  {http://meetings.aps.org/link/BAPS.2017.MAR.A27.6} {\bibfield  {journal}
  {\bibinfo  {journal} {http://meetings.aps.org/link/BAPS.2017.MAR.A27.6}\ }
  (\bibinfo {year} {2017}{\natexlab{a}})}\BibitemShut {NoStop}%
\bibitem [{\citenamefont {Jain}\ and\ \citenamefont
  {Kamilla}(1997)}]{JainKamilla}%
  \BibitemOpen
  \bibfield  {author} {\bibinfo {author} {\bibfnamefont {J.~K.}\ \bibnamefont
  {Jain}}\ and\ \bibinfo {author} {\bibfnamefont {R.~K.}\ \bibnamefont
  {Kamilla}},\ }\href@noop {} {\bibfield  {journal} {\bibinfo  {journal} {Phys.
  Rev. B.}\ }\textbf {\bibinfo {volume} {55}},\ \bibinfo {pages} {R4895(R)}
  (\bibinfo {year} {1997})}\BibitemShut {NoStop}%
\bibitem [{\citenamefont {Shao}\ \emph {et~al.}(2015)\citenamefont {Shao},
  \citenamefont {Kim}, \citenamefont {Haldane},\ and\ \citenamefont
  {Rezayi}}]{Shao}%
  \BibitemOpen
  \bibfield  {author} {\bibinfo {author} {\bibfnamefont {J.}~\bibnamefont
  {Shao}}, \bibinfo {author} {\bibfnamefont {E.-A.}\ \bibnamefont {Kim}},
  \bibinfo {author} {\bibfnamefont {F.~D.~M.}\ \bibnamefont {Haldane}}, \ and\
  \bibinfo {author} {\bibfnamefont {E.~H.}\ \bibnamefont {Rezayi}},\ }\href
  {\doibase 10.1103/PhysRevLett.114.206402} {\bibfield  {journal} {\bibinfo
  {journal} {Phys. Rev. Lett.}\ }\textbf {\bibinfo {volume} {114}},\ \bibinfo
  {pages} {206402} (\bibinfo {year} {2015})}\BibitemShut {NoStop}%
\bibitem [{\citenamefont {Wang}\ \emph
  {et~al.}(2017{\natexlab{b}})\citenamefont {Wang}, \citenamefont {Geraedts},
  \citenamefont {Rezayi},\ and\ \citenamefont {Haldane}}]{Jie_MonteCarlo}%
  \BibitemOpen
  \bibfield  {author} {\bibinfo {author} {\bibfnamefont {J.}~\bibnamefont
  {Wang}}, \bibinfo {author} {\bibfnamefont {S.}~\bibnamefont {Geraedts}},
  \bibinfo {author} {\bibfnamefont {E.~H.}\ \bibnamefont {Rezayi}}, \ and\
  \bibinfo {author} {\bibfnamefont {F.~D.~M.}\ \bibnamefont {Haldane}},\
  }\href@noop {} {\bibfield  {journal} {\bibinfo  {journal} {arXiv:cond-mat}\
  ,\ \bibinfo {pages} {1710.09729}} (\bibinfo {year}
  {2017}{\natexlab{b}})}\BibitemShut {NoStop}%
\bibitem [{Note1()}]{Note1}%
  \BibitemOpen
  \bibinfo {note} {A downside to writing the wavefunction in terms of the
  $f(z)$ is that it does not look holomorphic. However, the factor of $\protect
  \qopname \relax o{exp}(-\protect \frac {1}{2mN}zz^*)$ contained in the
  definition of $f(z)$ does cancel the $z^*$ which appear in this
  equation}\BibitemShut {NoStop}%
\bibitem [{\citenamefont {Girvin}(1984)}]{GirvinPH}%
  \BibitemOpen
  \bibfield  {author} {\bibinfo {author} {\bibfnamefont {S.~M.}\ \bibnamefont
  {Girvin}},\ }\href {\doibase 10.1103/PhysRevB.29.6012} {\bibfield  {journal}
  {\bibinfo  {journal} {Phys. Rev. B}\ }\textbf {\bibinfo {volume} {29}},\
  \bibinfo {pages} {6012} (\bibinfo {year} {1984})}\BibitemShut {NoStop}%
\bibitem [{\citenamefont {Wang}\ and\ \citenamefont
  {Senthil}(2016)}]{WangSenthil}%
  \BibitemOpen
  \bibfield  {author} {\bibinfo {author} {\bibfnamefont {C.}~\bibnamefont
  {Wang}}\ and\ \bibinfo {author} {\bibfnamefont {T.}~\bibnamefont {Senthil}},\
  }\href {\doibase 10.1103/PhysRevB.93.085110} {\bibfield  {journal} {\bibinfo
  {journal} {Phys. Rev. B}\ }\textbf {\bibinfo {volume} {93}},\ \bibinfo
  {pages} {085110} (\bibinfo {year} {2016})}\BibitemShut {NoStop}%
\end{thebibliography}%
\end{document}